\documentclass[prl, floatfix, nobibnotes, reprint, superscriptaddress]{revtex4-1}

\usepackage[english]{babel}
\usepackage[utf8]{inputenc}
\usepackage[T1]{fontenc}
\usepackage{amsmath}
\usepackage{mathalpha}
\usepackage{amssymb}
\usepackage{amsfonts}
\usepackage{graphicx}
\usepackage{relsize}
\usepackage{bm}% bold math
\usepackage{multirow}
\usepackage{bigints}
\usepackage[varg]{txfonts}
\usepackage{bbm}
\usepackage{listings}
\usepackage[hmargin=2.5cm, vmargin=4cm]{geometry}
\graphicspath{{figures/}}
\usepackage{url}
\usepackage{braket}
\usepackage{textcomp}
\usepackage{float}
\usepackage{bbold} %doppelgestrichene Zeichen möglich
\usepackage{comment}   % For the comment enviroment
\usepackage{suffix}
\usepackage[dvipsnames]{xcolor} %\textcolor{Red}{}
\usepackage[normalem]{ulem}

\newcommand{\papertitle}{The Alchemical Integral Transform revisited}
\DeclareMathOperator{\diag}{diag}
\allowdisplaybreaks

\begin{document}

\title{\papertitle}

\author{Simon León Krug}
%\email{simonleon.krug@univie.ac.at}
\affiliation{Machine Learning Group, Technische Universit\"at Berlin, 10587 Berlin, Germany}
%\affiliation{University of Vienna, Vienna Doctoral School in Physics, Boltzmanngasse 5, 1090 Vienna, Austria}

\author{O. Anatole von Lilienfeld}
\email{anatole.vonlilienfeld@utoronto.ca}
\affiliation{Machine Learning Group, Technische Universit\"at Berlin, 10587 Berlin, Germany}
\affiliation{Berlin Institute for the Foundations of Learning and Data, 10587 Berlin, Germany}
\affiliation{Chemical Physics Theory Group, Department of Chemistry, University of Toronto, St. George Campus, Toronto, ON, Canada}
\affiliation{Department of Materials Science and Engineering, University of Toronto, St. George Campus, Toronto, ON, Canada}
% \affiliation{Vector Institute for Artificial Intelligence, Toronto, ON, M5S 1M1, Canada}
\affiliation{Vector Institute for Artificial Intelligence, Toronto, ON, Canada}
\affiliation{Department of Physics, University of Toronto, St. George Campus, Toronto, ON, Canada}
\affiliation{Acceleration Consortium, University of Toronto, Toronto, ON, Canada}

\date{\today}

\begin{abstract}
%Within quantum alchemy, observables of one system~$B$, such as the energy~$E_B$, can be obtained by perturbing the external potential of a different iso-electronic system~$A$, by Taylor expansion of~$E_A$ in an alchemical change~$\lambda$. 
We recently introduced the Alchemical Integral Transform (AIT) enabling the prediction of energy differences, and guessed an Ansatz to parametrize space~$\bm{r}$ in some alchemical change~$\lambda$. 
Here, we present a rigorous derivation of AIT's kernel~$\mathcal{K}$ and discuss the parametrization~$\bm{r}(\lambda)$ in $n$ dimensions, i.e. necessary conditions, mathematical freedoms and additional constraints when obtaining it. Analytical expressions for changes in energy spectra and densities are given for a number of systems. Examples include homogeneous potentials like the quantum harmonic oscillator, Hydrogen-like atom, and Dirac well, both for one- and multiparticle cases, and a multiparticle system beyond coordinate scaling for harmonic potentials.
\end{abstract}

\maketitle

\section{Introduction}
Schrödinger's equation yields the absolute energy spectrum and corresponding eigen states. However, most, if not all, processes of interest in chemistry and materials science deal only with relative changes between systems $A$ and $B$. %\cite{beste_harrison_2006}.
Early relative computations, e.g. treating nuclear charges or entire functional groups as (non-discrete) parameters, trace back to Hückel~\cite{hueckel_1931},  Hylleraas \& Midtal~\cite{hylleraas_1956}, Wilson \cite{bright_wilson}, Politzer \& Parr~\cite{politzer_parr_1974} and Levy \cite{levy_1978,levy_1979}.
In this sense, nuclear transmutations aka computational alchemy, simply correspond to yet another fundamental method of inferring information from one quantum system to another. 
The extensive work done on 1D systems and their application to higher dimensions is dedicated to such fundamental questions ~\cite{one_dim_mimicking,takahashi2005thermodynamics,mattis1993manybody,schlottmann1997exact_in1D,busch2003low,magyar2007ground}. %, or more recently, in the prediction of bonding properties from models like the quantum drude oscillator (https://doi.org/10.48550/arXiv.2309.14910), which is again well-known Quantum Harmonic Oscillator (QHO),
Another example comes from conceptual DFT~\cite{CDFT,reactivity_CDFT,FUENTEALBA,quintana_2022,gazquez_2021}.

Alchemical approaches may realize such inference with rigor in that only parameters are changed, instead of adding electrons or entire dimensions.
Some of the more recent applications of alchemy to quantum mechanical problems include the exploration of chemical compound space~\cite{chemical_space_balawender,shiraogawa_2022}, design of large band-gap (III)-(IV) semi-conductors~\cite{Samuel2018bandgaps}, 
treatment of alchemically symmetric molecules~\cite{von_Rudorff_2020,rudorff2020_alchem_chirality,vonlilienfeld_2023}, reactions like deprotonations~\cite{vonrudorff_2020_rapid,munoz_2020}, bonds~\cite{Samuel-JCP2016,eikey_2022_diatomic}, or excitations~\cite{eikey_2022_excitations}.

In a previous paper~\cite{krug_DeltaE}, we studied relative energies using an Alchemical Integral Transform (AIT), allowing its user to fully recover the energy of a final system~$E_B$ from an iso-electronic initial system's electron density~$\rho_A$ and energy~$E_A$. 
AIT's utility hinges on finding a suitable parametrization $\bm{r}(\lambda)$, with $\lambda$ being the parameter of alchemical change. 
The need for the parametrization emerged %from the original definition of the problem: consider two iso-electronic systems, $\hat{H}_A$ and $\hat{H}_B$, connected by an interpolation parameter $\lambda$ such that $\hat{H}(\lambda) = (1-\lambda)\hat{H}_A + \lambda \hat{H}_B $. Then the corresponding electron density is $\rho(\lambda, \bm{r})$. $\bm{r}(\lambda)$~emanated
from rewriting a general electron density~$\rho(\lambda, \bm{y})$ such that any $\lambda$-dependency rested with a parametrization~$\bm{r}$ of the coordinates (cf.~Eqs.~\ref{eq:HellmannFeynmanApplied} and~\ref{eq:ansatz_rho}), and  $\bm{r}$~was found by trial and error. 
%This is not just an unsatisfactory \textit{modus operandi} in general, but it leaves us without any systematic tool to gauge the quality of our assertion in specific.

Here, we present a rigorous and compact derivation of AIT's kernel $\mathcal{K}$, and discuss conditions, constraints and mathematical freedoms of finding~$\bm{r}(\lambda)$ in $n$~dimensions. First, we consider the constraint of probability conservation: then, $\bm{r}(\lambda)$ can be found as long as the coordinates of systems $A,B$ can be expressed as one another by an affine transformation, i.e. $\bm{x} \rightarrow A\bm{x} + b$ where $A\in \mathbb{R}^n \times \mathbb{R}^n$ is an invertible matrix and $\bm{b} \in \mathbb{R}^n$.
Furthermore, we look at the constraint of a known electric dipole moment.

By extension, we obtain analytical expressions of relative electron densities, the functional behavior of energies with respect to the system's parameters, and conservation laws for the electron densities. 
Examples include homogeneous potentials (iso-tropic quantum harmonic oscillator, hydrogen-like atom, Dirac well) in one- and multiparticle cases and a fictitious harmonic multiparticle system.

% \section{Theory}

\section{A new derivation}

Consider any two iso-electronic systems with electronic Hamiltonians $\hat{H}_A$ and $\hat{H}_B$ and their external potentials $v_A$ and $v_B$. Connect them via a linear transformation $\hat{H}(\lambda) := \hat{H}_A (1-\lambda)+ \hat{H}_B \lambda$ such that we obtain a general electron density $\rho(\lambda, \bm{y})$ at every point $\lambda \in [0,1]$ with $\rho_A := \rho(\lambda = 0)$, $\rho_B := \rho(\lambda = 1)$.
The first-order derivative of the general energy according to the Hellmann-Feynman theorem corresponds to~\cite{rudorff2020_alchem_chirality}:
\begin{align}
	\frac{\partial E(\lambda)}{\partial \lambda} &= \bra{\Psi_{\lambda}}\hat{H}_B - \hat{H}_A\ket{\Psi_{\lambda}}\\
    \label{eq:HellmannFeynmanApplied}
	&=\int_{\mathbb{R}^n} d\bm{y}\,\, \Delta v(\bm{y}) \,\, \rho(\lambda, \bm{y})
\end{align}
\noindent
with potential difference $\Delta v(\bm{y}) = v_B(\bm{y}) - v_A(\bm{y})$ and general electron density $\rho(\lambda, \bm{y})$ along~$\lambda$.
% Note that AIT is general, and not limited to Coulombic systems (i.e. molecular Hamiltonians). Rather it extends to any Schrödinger equation, i.e.~$\rho$ could refer to any probability density, not just the electronic one.

The original derivation in Ref.~\onlinecite{krug_DeltaE} continued with a perturbative expansion of the energy in $\lambda$; here, we employ the converse approach with thermodynamic integration~\cite{von_Rudorff_2020}:
\begin{align}
    \Delta E := & \, E_B - E_A \\
    \label{eq:therm_interpolation}
    = & \int_0^1 d\lambda \, \int_{\mathbb{R}^n} d\bm{y}\,\, \Delta v(\bm{y}) \,\, \rho(\lambda, \bm{y})
\end{align}
Now, we rewrite the general electron density $\rho(\lambda, \bm{y})$ in terms of the initial electron density $\rho_A$ and the parametrization~$\bm{r}(\lambda)$ (which implicitly depends on~$\bm{y}$):
\begin{align}
    \label{eq:ansatz_rho}
    \rho(\lambda, \bm{y}) = \mathcal{N}(\lambda) \, \rho_A (\bm{r}(\lambda) )
\end{align}
This differs from the original definition~\cite{krug_DeltaE} in that we introduced a normalization $\mathcal{N}(\lambda)$. In doing so, $\bm{r}(\lambda)$~now only needs to transform the coordinates of system $A$ such that the \textit{functional form} of the intermediate system is ensured. 

Inserting Eq.~\ref{eq:ansatz_rho} into Eq.~\ref{eq:therm_interpolation} and reordering, we find a general kernel of AIT, dubbed~$\mathcal{K}[\Delta v](\bm{x})$:
\begin{align}
    \Delta E
    &= \!\int_0^1 d\lambda \int_{\mathbb{R}^n}\!\! d\bm{y}\,\Delta v(\bm{y}) \, \mathcal{N}(\lambda) \, \rho_A (\bm{r}(\lambda) )\\
    &= \!\int_0^1 d\lambda \int_{\mathbb{R}^n} \!\!d\bm{y}\,\Delta v(\bm{y}) \, \mathcal{N}(\lambda) \!\! \int_{\mathbb{R}^n} \!\! d\bm{x}\, \rho_A (\bm{x}) \, \delta^{(n)} (\bm{x} - \bm{r}(\lambda) ) \\
    \label{eq:general_K}
    &= \!\int_{\mathbb{R}^n} \!\! d\bm{x}\, \rho_A (\bm{x}) \!\!\underbrace{\int_0^1 d\lambda  \, \mathcal{N}(\lambda) \!\! \int_{\mathbb{R}^n} \!\!d\bm{y}\,\Delta v(\bm{y}) \, \delta^{(n)} (\bm{x} - \bm{r}(\lambda) )}_{:= \mathcal{K}[\Delta v](\bm{x})}
\end{align}
Thus, we rephrased the problem's original question from "What are the eigen values and eigen functions of the Hamiltonian?" to "What is the mathematical structure of the coordinates of the underlying Hamiltonian?". Both questions are equivalent but the latter is more accessible as we will show below.

\section{The parametrization $\bm{r}(\lambda)$}

To progress, we want to find a solution to Eq.~\ref{eq:ansatz_rho} without solving for the density directly.
Using constraints like the probability conservation of the density restricts AIT to problems where $\bm{r}(\lambda) = A(\lambda) \cdot \bm{y} + \bm{b}(\lambda)$ (with some invertible matix $A(\lambda)$ and a vector offset $\bm{b}(\lambda)$), i.e. systems whose coordinates can be expressed as one another by an affine transformation. With affine transformations, we can easily solve:
\begin{align}
    \rho(\lambda, \bm{y}) &= \mathcal{N}(\lambda) \, \rho_A (A(\lambda) \cdot \bm{y} + \bm{b}(\lambda)) \\
    \label{eq:solve_for_N}
    \Leftrightarrow \int_{\mathbb{R}^n} \!\! d\bm{y} \, \rho(\lambda, \bm{y}) &= \mathcal{N}(\lambda) \! \int_{\mathbb{R}^n} \!\! d\bm{y} \, \rho_A (A(\lambda) \cdot \bm{y} + \bm{b}(\lambda))
\end{align}
The left-hand side is just the total number of particles along the alchemical path $\lambda$, which remains a constant~$N$. The integral on the right-hand side can be solved by substituting $\bm{r} = A(\lambda) \cdot \bm{y} + \bm{b}(\lambda)$ with a Jacobian of $ |\det \left( \partial \bm{r}/ \partial \bm{y} \right)| = | \det (A(\lambda)) |$.
\begin{align}
    \Rightarrow \quad N &= \mathcal{N}(\lambda) \int_{\mathbb{R}^n} \!\! d\bm{u} \, \frac{\rho_A (\bm{u})}{| \det (A(\lambda)) |} \\
    \label{eq:N_of_lambda}
    \Leftrightarrow \quad \mathcal{N}(\lambda) &=  | \det (A(\lambda)) |
\end{align}
Inserting this into $\mathcal{K}[\Delta v](\bm{x})$ as defined in Eq.~\ref{eq:general_K}:
\begin{widetext}
    \begin{align}
        \mathcal{K}[\Delta v](\bm{x}) := &\int_0^1 d\lambda  \, | \det (A(\lambda)) | \int_{\mathbb{R}^n} d\bm{y}\,\Delta v(\bm{y}) \, \delta^{(n)} (\bm{x} - A(\lambda) \cdot \bm{y} - \bm{b}(\lambda) ) \\
        = &\int_0^1 d\lambda  \, | \det (A(\lambda)) | \int_{\mathbb{R}^n} d\bm{y}\,\Delta v(\bm{y}) \, \delta^{(n)} (A(\lambda) \cdot [A^{-1}(\lambda) \, (\bm{x} - \bm{b}(\lambda)) - \bm{y}]) \\
        = &\int_0^1 d\lambda  \,  \int_{\mathbb{R}^n} d\bm{y}\,\Delta v(\bm{y}) \, \delta^{(n)} (A^{-1}(\lambda) \, (\bm{x} - \bm{b}(\lambda)) - \bm{y}) \\
        \label{eq:K_final_form}
        =& \int_0^1 d\lambda  \,\, \Delta v \left( A^{-1}(\lambda) \, (\bm{x} - \bm{b}(\lambda)) \right)
    \end{align}    
\end{widetext}

However, the parametrization~$\bm{r}(\lambda)$ is not limited to affine transformations of the coordinates. Eq.~\ref{eq:ansatz_rho} allows to include multiple constraints beyond probability conservation. In fact, this is one advantage of AIT since materials design is generally not interested in arbitrary iso-electronic energy differences, but rather those which are subject to constraints. 

For example, consider two systems whose first component of the electric dipole moment is known at coordinate component~$b_1(\lambda)$ along~$\lambda$:
\begin{align}
    \label{eq:electric_dipole_moment}
    p_1 (b_1(\lambda)) &= \int_{\mathbb{R}^n} \!\! d\bm{y} \, (y_1 - b_1(\lambda))\, \rho_A (\bm{y}) = \text{const.}
\end{align}
This constraint allows for a different parametrization:
\begin{align}
    \label{eq:param_dipole}
    \bm{r}(\lambda) &= 
    \begin{pmatrix}
        \sqrt{2|y_1|} + b_1(\lambda) \\
        A'(\lambda) \bm{y}' + \bm{b}'(\lambda)
    \end{pmatrix}
\end{align}
with primed quantities excluding the first component, i.e. invertible~$(n-1) \times (n-1)$-matrix~$A'(\lambda)$, coordinate vector $\bm{y}' = (y_2, \dots, y_n)^T$ and vector offset~$\bm{b}'(\lambda) = (b_2(\lambda), \dots, b_n(\lambda))^T$.
Inserting Eq.~\ref{eq:param_dipole} into Eq.~\ref{eq:ansatz_rho}, we find:
\begin{align}
    \int_{\mathbb{R}^n} \!\! d\bm{y} \, \rho(\lambda, \bm{y}) &= \mathcal{N}(\lambda) \! \int_{\mathbb{R}^n} \!\! d\bm{y} \, \rho_A (\bm{r}(\lambda)) \\
    \Leftrightarrow \quad N &= \mathcal{N}(\lambda) \! \int_{\mathbb{R}^n} \!\! d\bm{r} \, \frac{\rho_A (\bm{r})}{\big\vert \det \left( \frac{\partial \bm{r}}{\partial \bm{y}} \right) \big\vert} \\
    &= \mathcal{N}(\lambda) \! \int_{\mathbb{R}^n} \!\! d\bm{r} \, \frac{|r_1 - b_1(\lambda)|}{| \det \left( A'(\lambda) \right)|} \, \rho_A (\bm{r}) \\
    \Leftrightarrow \quad \mathcal{N}(\lambda) &= \frac{N}{|p_1|} \, | \det \left( A'(\lambda) \right)|
\end{align}

With knowledge of~$\mathcal{N}(\lambda)$, we can calculate~$\mathcal{K}[\Delta v](\bm{x})$ as defined in Eq.~\ref{eq:general_K}:
\begin{widetext}
    \begin{align}
        \mathcal{K}[\Delta v](\bm{x}) := &\int_0^1 d\lambda  \, \frac{N}{|p_1|} \, | \det \left( A'(\lambda) \right)| \int_{\mathbb{R}^n} d\bm{y}\,\Delta v(\bm{y}) \, \delta^{(n)} (\bm{x} - \bm{r}(\lambda) ) \\
        = & \int_0^1 d\lambda  \, \frac{N}{|p_1|} \, \int_{\mathbb{R}^n} d\bm{y}\,\Delta v(\bm{y}) \,|x_1 + b_1(\lambda)| \,\delta \left( y_1 - \frac{(x_1 + b_1(\lambda))^2}{2} \right) | \det \left( A'(\lambda) \right)| \, \delta^{(n-1)}\left(\bm{x}' - A'(\lambda) \bm{y}' - \bm{b}'(\lambda) \right) \\
        = & \int_0^1 d\lambda  \, \frac{N}{|p_1|} \, \int_{\mathbb{R}^n} d\bm{y}\,\Delta v(\bm{y}) \,|x_1 + b_1(\lambda)| \,\delta \left( y_1 - \frac{(x_1 + b_1(\lambda))^2}{2} \right) \, \delta^{(n-1)}\left( A'^{-1}(\lambda) (\bm{x}' - \bm{b}'(\lambda)) - \bm{y}' \right)\\
        = & \, \frac{N}{|p_1|} \int_0^1 d\lambda \,|x_1 + b_1(\lambda)| \, \Delta v(\bm{y}) \left( \frac{(x_1 + b_1(\lambda))^2}{2}, \,A'^{-1}(\lambda) (\bm{x}' - \bm{b}'(\lambda)) \right)
    \end{align}    
\end{widetext}

However, even without any additional constraints the parametrization~$\bm{r}(\lambda)$ of the coordinates can be extended by a function~$\tilde{r}_i (||\bm{y}||_2)$ in dimensions $n> 1$ due to the behavior of the determinant:
\begin{align}
    \det \left( \frac{\partial \tilde{r}_i}{\partial y_k} \right) = \frac{1}{||\bm{y}||_2} \det \Bigg( \underbrace{ \frac{\partial \tilde{r}_i}{\partial ||\bm{y}||_2} \, y_k }_{= \bm{\nabla} \bm{\tilde{r}} \, \otimes \, \bm{y}}\Bigg)
\end{align}
$\bm{\nabla} \bm{\tilde{r}} \, \otimes \, \bm{y}$ is the outer product of~$\bm{\nabla} \bm{\tilde{r}}$ and~$\bm{y}$ and thus, has rank~1. But the determinant of any matrix with no full rank is always zero by the invertible matrix theorem such that
\begin{align}
    \det \left( \frac{\partial \tilde{\bm{r}}}{\partial \bm{y}} \right) = 0
\end{align}
for any dimension greater 1. Consequently, all parametrizations which depend only on $||\bm{y}||_2$ may be included.

Further constraints of the problem allow to extend the possible parametrizations~$\bm{r}(\lambda)$. This method of constraints implicates that $\mathcal{K}[\Delta v](\bm{x})$ can be found just from the problem's statement in the Hamiltonian.% Finding an affine transformation, or an~$\bm{r}(\lambda)$ like Eq.~\ref{eq:param_dipole}, allows analytical statements about the behavior of the energy difference.

Note how Eq.~\ref{eq:ansatz_rho} made no assumption about excited states whatsoever; in fact, it conserves the excitation between initial and final system.

\section{The final density}
A corollary from the derivation of transformations is a direct statement about the density of the final system~$\rho_B$. For example, since we assumed an affine transformation in Eq.~\ref{eq:ansatz_rho} and then proceeded to compute $\mathcal{N}(\lambda)$ in Eq.~\ref{eq:N_of_lambda}, we immediately find:
\begin{align}
    \label{eq:electron_density_corollary}
    \rho_B (\bm{x}) = | \det (A(1)) | \, \, \rho_A \bigg( A(1) \cdot \bm{x} + \bm{b}( 1) \bigg)
\end{align}
where $A(\lambda = 1)$ and $\bm{b}(\lambda = 1)$ are a constant matrix and vector, respectively. This means, if two Hamiltonians are related via an affine transformation of their coordinates (disregarding any normalization) then their electron densities are related by Eq.~\ref{eq:electron_density_corollary} and only the final configuration~$A(\lambda = 1), \bm{b}(\lambda = 1)$ is necessary to connect them.

A similar statement can be made for systems with known electric dipole moment~$p_1$:
\begin{align}
    \notag
    \rho_B (\bm{x}) = &\, \frac{N}{|p_1|} \, | \det \left( A'(1) \right)| \\ 
    &\quad \times \rho_A \bigg( \sqrt{2|x_1|} + b_1(1), \, A'(1) \bm{x}' + \bm{b}'(1) \bigg)
    \label{eq:electron_density_corollary_dipole}
\end{align}

% \section{Results}
% \label{sec:results}

\section{Homogeneous potentials}

Let us consider a class of examples: consider this $n$-dimensional Schrödinger equation:% (and use multi-index notation $\bm{y}^{\bm{\alpha}} := y_1^{\alpha_1}y_2^{\alpha_2} \dots$, $|\bm{\alpha}| := \alpha_1 + \alpha_2 + \dots$):
\begin{align}
    \label{eq:homfunc_Hamiltonian}
    \hat{H}^{\text{hom}} := & -\frac{1}{2} \bm{\nabla}^2_{\bm{y}} + k_A^2 \, f(k_A \bm{y})
\end{align}
where $f$ is any (positive) homogeneous function of degree $\nu \neq -2$, i.e. $f(k_A\bm{y}) = k_A^{\nu} \, f(\bm{y})$, and $k_A$ is a real, positive constant describing system~$A$. 
Transforming this Hamiltonian at parameter~$k_A$ into one at parameter~$k(\lambda)$ necessiates a scaling transformation of the coordinates, $\bm{y} \rightarrow k(\lambda)/k_A \, \bm{y}$:
\begin{align}
    \label{eq:homfunc_Hamiltonian_rescaled}
    \hat{H}^{\text{hom}} := & k_A^2 \left( -\frac{1}{2} \frac{\bm{\nabla}^2_{\bm{y}}}{k^2 (\lambda)} + f(k (\lambda) \bm{y}) \right)
\end{align}
The prefactor of $k_A^2$ is accounted for by~$\mathcal{N}(\lambda)$.

Clearly, this is a restriction of affine transformations to $A(\lambda) = k(\lambda)/k_A \, \mathbb{1}_n$ ($\mathbb{1}_n$ representing identity in $n$ dimensions) and $b(\lambda) = 0$. Because of this, degree $\nu = -2$ must be excluded as it would constitute the case where Schrödinger's equation includes no parameter to be changed and AIT does not apply.

\begin{figure}
    \centering
    \includegraphics[width=\linewidth]{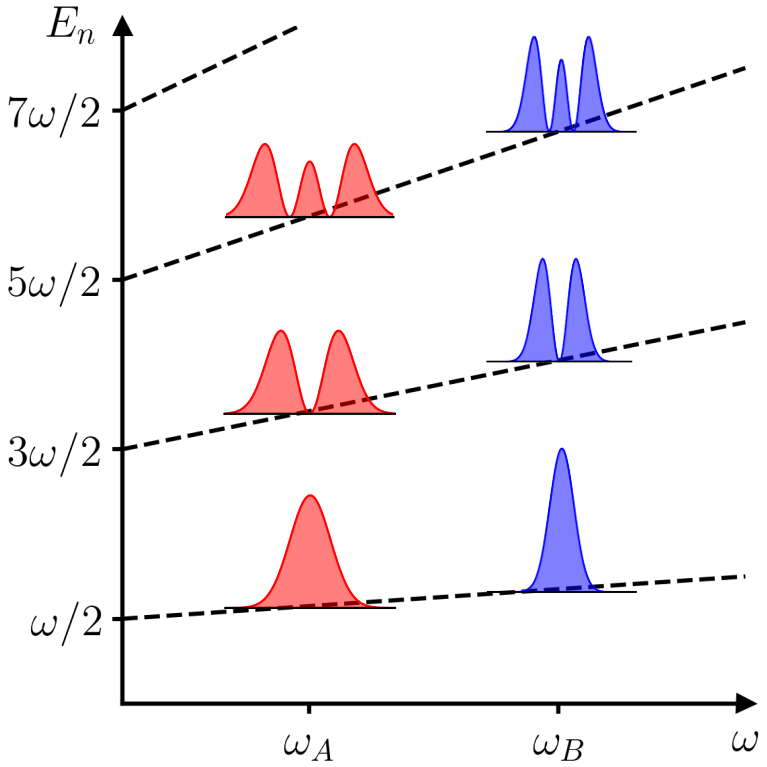}
    \caption{Visual representation of spectrum and density rescaling derived from AIT in case of the quantum harmonic oscialltor (QHO) at different excitations.}
    \label{fig:QHO_cartoon}
\end{figure}

To calculate the kernel~$\mathcal{K}$ for energy differences between system~$A$ and~$B$, we also need the potential difference $\Delta v(\bm{y}) = (k^{2+\nu}_B - k^{2+\nu}_A) f(\bm{y})$ and $k(\lambda)$:
\begin{align}
    k^{2+\nu}(\lambda) f(\bm{x}) &= (1 - \lambda) k_A^{2+\nu} f(\bm{x}) + \lambda k_B^{2+\nu} f(\bm{x}) \\
    \Leftrightarrow \quad k(\lambda) &= \left[(1 - \lambda) k_A^{2+\nu} + \lambda k_B^{2+\nu} \right]^{1/(2 + \nu)} \\
    \intertext{With this knowledge, we find:}
    \Rightarrow \quad \mathcal{K}[\Delta v](\bm{x}) &= \int_0^1 d\lambda  \,\, \Delta v \left( A^{-1}(\lambda) \bm{x} \right) \\
    &= (k_B^{2+\nu} - k_A^{2+\nu}) \int_0^1 \! d\lambda \,\, f\left(\frac{k_A}{k(\lambda)} \bm{x} \right) \\
    &= \int_0^1 \!\! d\lambda  \frac{ (k_B^{2+\nu} - k_A^{2+\nu}) \, f\left(k_A \bm{x} \right)}{\left[ (1 - \lambda) k_A^{2+\nu} + \lambda k_B^{2+\nu} \right]^{\nu / (2 + \nu)}}  \\
    \label{eq:kernel_hom}
    % &= k_A^2 \, f\left(k_A \bm{x} \right) \frac{2 + \nu}{2} \left( \frac{k_B^2}{k_A^2} - 1 \right) \\
    &= f\left( \bm{x} \right) \frac{2 + \nu}{2} \left( k_B^2 - k_A^2 \right) \, k_A^{\nu}
\end{align}
The corresponding relative density (cf. Eq.~\ref{eq:electron_density_corollary}) is found to be:
\begin{align}
    \label{eq:relation_hom_densities}
    \rho^{\text{hom}}_B(\bm{x}) &=  \left(\frac{k_B}{k_A} \right)^{n} \,\,\rho_A^{\text{hom}}\left( \frac{k_B}{k_A} \bm{x} \right)
\end{align}
Special cases are
\begin{itemize}
    \item the quantum harmonic oscillator with $f(\bm{y}) = y^2/2$, $k_{A,B} = \sqrt{\omega_{A,B}}$, $\nu = 2$:
    \begin{align}
        \label{eq:kernel_QHO}
        \mathcal{K}(x) &= (\omega_B - \omega_A)\omega_A x^2 \\
        \Rightarrow \Delta E &= \int_{\mathbb{R}} dx \, \rho_A^{\text{QHO}} \, \mathcal{K}(x) \\
        \label{eq:integral_QHO}
        &= (\omega_B - \omega_A)\omega_A \int_{\mathbb{R}} dx \,x^2 \, \rho_A^{\text{QHO}} \\
        &= (\omega_B - \omega_A)\left(n + \frac{1}{2} \right) \\
        \label{eq:relation_QHO_densities}
        \rho^{\text{QHO}}_B(x) &= \sqrt{\omega_B / \omega_A}  \,\,\rho_A^{\text{QHO}}(\sqrt{\omega_B / \omega_A} \, x)
    \end{align}
    A visualization can be found in Fig.~\ref{fig:QHO_cartoon}. 
    The $n$-dimensional isotropic quantum harmonic oscillator (QHO) works analogously.
    
    \item the Hydrogen-like atom with $f(\bm{y}) = -1/||\bm{y}||_2$, $k_{A,B} = Z_{A,B}$, $\nu = -1$:
    \begin{align}
        \label{eq:kernel_hydlike}
        \mathcal{K}(\bm{x}) &= \frac{-Z_B+Z_A}{2||\bm{x}||_2} \left(1 + \frac{Z_B}{Z_A} \right) \\
        \Rightarrow \Delta E &= \int_{\mathbb{R}^3} d\bm{x} \, \rho_A^{\text{HL}} \, \mathcal{K}(\bm{x}) \\
        \label{eq:integral_HL}
        &= \frac{-Z_B+Z_A}{2} \left(1 + \frac{Z_B}{Z_A} \right)\int_{\mathbb{R}^3} d\bm{x} \, \frac{\rho_A^{\text{HL}}}{||\bm{x}||_2} \\
        &= \frac{-Z_B^2+Z_A^2}{2n^2} \\
        \label{eq:relation_HL_densities}
        \rho^{\text{HL}}_B(\bm{x}) &= \left( Z_B/Z_A \right)^3 \,\rho_A^{\text{HL}}(Z_B/Z_A \, \bm{x})
    \end{align}

    \item the Dirac well of depth $d$ with $f(\bm{y}) = -\delta(y)$, $k_{A,B} = d_{A,B}$, $\nu = -1$:
    \begin{align}
        \label{eq:kernel_Dirac}
        \mathcal{K}(x) &= \frac{d_A^2-d_B^2}{2d_A} \delta(x) \\
        \Rightarrow \Delta E &= \int_{\mathbb{R}} dx \, \rho_A^{\text{Dirac}} \, \mathcal{K}(x) \\
        &= \frac{d_A^2-d_B^2}{2d_A} \int_{\mathbb{R}} dx \, \delta(x) \, d_A\, e^{-2d_A \, |x|} \\
        &= \frac{d_A^2-d_B^2}{2} \\
        \rho^{\text{Dirac}}_B(x) &= (d_B / d_A )  \,\,\rho_A^{\text{Dirac}}((d_B / d_A )  \, x)
    \end{align}
\end{itemize}
The explicit evaluation of the integrals in Eqs.~\ref{eq:integral_QHO} and~\ref{eq:integral_HL} and the explicit electron densities can be found in the~SI~\cite{cohentannoudji}. AIT~correctly reproduces the well-known results in all three cases above. A generalization to quasi-homogeneous functions is straightforward.

\section{Multi-particle systems}
\label{sec:multi-particle}

In multielectron systems, it is no longer obvious how the parametrization~$\bm{r}(\lambda)$, a transform of $\bm{y}$, can be found, due to multiple coordinates $\bm{y}_i$ for each electron. 
This is quickly resolved via the definition of $\rho(\bm{y})$~\cite{szabo_ostlund}. Consider an $N$-electron problem:
\begin{align}
    \rho(\bm{y})\! :=& \! \sum_{i = 1}^{N} \bra{\Psi (\bm{y}_1, \dots, \bm{y}_{N})} \delta^{(n)} (\bm{y} - \bm{y}_i) \ket{\Psi (\bm{y}_1, \dots, \bm{y}_{N})}\\
    =& \, N  \bra{\Psi (\bm{y}_1, \dots, \bm{y}_{N})} \delta^{(n)} (\bm{y} - \bm{y}_1) \ket{\Psi (\bm{y}_1, \dots, \bm{y}_{N})}
\end{align}
The second line applies to indistinguishable, and thus interchangeable, electrons and hence, all coordinates~$\bm{y}_i$ must transform identically. % and hence~$\bm{r}(\lambda, \bm{y}) = \bm{r}(\lambda, \bm{y}_i)$.
% However, distinct particles may be transformed differently.
For an application to real multielectron systems, we refer to Ref.~\onlinecite{krug_2024_atomic} (and specifically the Hamiltonian~$\hat{H}^{\text{fic}}$). Due to the nature of such systems, Ref.~\onlinecite{krug_2024_atomic} must resort to comparisons with numerical and experimental results, while in this work we will present only a theoretical basis and two examples. 
As we might consider interacting particles which behave not necessarily like electrons, we will instead referring to $\rho$ as a (particle) density from here on. 

Consider the homogeneous system presented in Eq.~\ref{eq:homfunc_Hamiltonian} but now with $N$ particles and interparticle repulsion proportional to distance$^{-2}$:
\begin{align}
    \label{eq:homfunc_multi_Hamiltonian}
    \hat{H}^{N,\text{hom}} := & \sum_{i = 1}^N \left( -\frac{1}{2} \bm{\nabla}^2_{\bm{y}_i} + k_A^2 \, f(k_A \bm{y}_i) \right) + \sum_{\substack{i,j=1\\i \neq j}}^N \frac{1}{|| \bm{y}_i - \bm{y}_j ||^2_2}
\end{align}
It is easy to check how this additional level of complexity leaves the derivation of Eqs.~\ref{eq:kernel_hom} and~\ref{eq:relation_hom_densities} untouched. Consequently, Eq.~\ref{eq:homfunc_multi_Hamiltonian} results in the same kernel~$\mathcal{K}$ and density relationship as before.

In contrast to the single particle systems, the density of multiparticle systems is usually not known and the integral
\begin{align}
    \label{eq:AIT_setup}
    \Delta E = \int_{\mathbb{R}^n} d\bm{x} \, \rho_A (\bm{x}) \, \mathcal{K}(\bm{x})
\end{align}
cannot be evaluated explicitly. But even for systems with cumbersome or unknown solutions, knowledge about~$\rho_A$ is not necessary to extract statements using AIT. The exact (and possibly excited!) energy difference derived from the kernel in Eq.~\ref{eq:kernel_hom} can be used twice, once for the energy difference between systems~$A,B$, then between~$A,B'$, to give:
\begin{align}
    \notag
    & \quad E_{B'} - E_B\\
    \label{eq:double_difference}
    &= \int_{\mathbb{R}^n} d\bm{x} \, \rho_A^{N,\text{hom}} \,\bigg( \mathcal{K}[v_{B'} - v_A](\bm{x}) - \mathcal{K}[v_{B} - v_A](\bm{x}) \bigg) \\
    &= \int_{\mathbb{R}^n} d\bm{x} \, \rho_A^{N,\text{hom}} f(\bm{x}) \frac{2+\nu}{2} k_A^{\nu}\left( k_{B'}^2 -k_B^2 \right)
\end{align}
As $E_{B'} - E_B$ is independent of~$A$, so must be the right-hand side. Consequently
\begin{align}
    \label{eq:energy_propto}
    E &= \frac{2+\nu}{2} c k^2\\
    c &= \int_{\mathbb{R}^n} d\bm{x} \, \rho^{N,\text{hom}} f(k \bm{x}) \quad ,
\end{align}
with $c$ being constant with respect to $k$, i.e.
\begin{align}
    \frac{dc}{dk} &= 0 \quad .
\end{align}

Although these examples of homogeneous potentials above corroborate the validity of AIT, results for densities (and, by extension, energy differences) could have also been obtained via coordinate scaling as employed in DFT \cite{gould_2023}. Especially Eq.~\ref{eq:relation_hom_densities} has been previously derived \cite{levy_1991,nagy_1995,kuemmel_2008}. 
However, AIT allows for more complicated transformations and goes beyond (scalar) coordinate scaling which we will show now.

Consider the following multi-particle system inside a harmonic potential with a variation of the interparticle repulsion:
\begin{align}
    \label{eq:Hamiltonian_multiQHO}
    \hat{H}^{N,\text{QHO}} := \sum_{i=1}^{N} & \left(- \frac{\bm{\nabla}^2_{\bm{y}_i}}{2}  + \frac{\bm{y}_i^T \Omega^2_A \bm{y}_i}{2} \right) + 
    \sum_{\substack{i,j=1\\i\neq j}}^N \sum_{\mu = 1}^n  \frac{z_{\mu}}{(\bm{y}_i - \bm{y}_j)^2_{\mu}}
\end{align}
The potential strength is encoded in the diagonal matrix~$\Omega_A = \diag ((\omega_A)_1, \dots, (\omega_A)_n)$, the interparticle strength is~$z_{\mu}$. This interparticle repulsion is not proportional to distance$^{-2}$ but the component-wise coordinate difference$^{-2}$. % with components~$\mu$.
This multiparticle system allows a separate treatment for each dimension and is clearly not separable into $N$~one-particle systems (if~$N > 1$). $\Omega_A$~being an arbitrary diagonal matrix (except for scalar multiples of $\mathbb{1}_n$, i.e. $(\omega_A)_1 = (\omega_A)_2 = \dots$) suffices to move beyond coordinate scaling. Eq.~\ref{eq:Hamiltonian_multiQHO} is, however, separable into its $n$ dimensions which we exploit in the parametrization
\begin{align}
    \bm{r}_{\mu}(\lambda) &= \underbrace{\sqrt{\frac{(\omega (\lambda))_{\mu}}{(\omega_A)_{\mu}}}}_{= A_{\mu \mu }(\lambda)} \, y_{\mu}\\
    \intertext{with}
    \omega_{\mu}^2 (\lambda) &= ((\omega_B)^2_{\mu} - (\omega_A)^2_{\mu})\lambda + (\omega_A)^2_{\mu} \quad ,
\end{align}
similar to an isotropic QHO.
Note that this parametrization still allows for the cancellation of kinetic and repulsive terms, so~Eq.~\ref{eq:HellmannFeynmanApplied} applies.
Again, we need not worry about the normalization although in this case, it is harder to see why: the definition of $\mathcal{N}(\lambda)$ in Eq.~\ref{eq:ansatz_rho} allows for a factorization into $n$ normalization constants~$\mathcal{N}_1(\lambda), \dots, \mathcal{N}_n (\lambda)$. Consequently, we could have written~$\hat{H}^{N,\text{QHO}}$ as a sum over $n$~dimensions and treated every dimension in Eq.~\ref{eq:ansatz_rho} separately.

Equipped with a parametrization, we find the kernel:
\begin{align}
    \mathcal{K}[\Delta v](\bm{x}) =& \int_0^1 d\lambda  \,\, \Delta v \left( A^{-1}(\lambda) \, \bm{x} \right)\\
    =& \int_0^1 d\lambda  \sum_{\mu = 1}^n \frac{x^2_{\mu}}{2\sqrt{((\omega_B)^2_{\mu} - (\omega_A)^2_{\mu})\lambda + (\omega_A)^2_{\mu}}} \\
    =& \sum_{\mu = 1}^n \left((\omega_B)_{\mu} - (\omega_A)_{\mu} \right) (\omega_A)_{\mu}x_{\mu}^2 \\
    =& \, \bm{x}^T \left( \Omega_B -  \Omega_A \right) \Omega_A \, \bm{x}
\end{align}
Write the energy difference and relative densities using AIT:
\begin{align}
    \Delta E &= \int_{\mathbb{R}^n} d\bm{x} \, \rho_A^{\text{multi}} \, \mathcal{K}(\bm{x}) \\ 
    \label{eq:energy_multi}
    &=  \int_{\mathbb{R}^n} d\bm{x} \, \rho_A^{\text{multi}} \, \sum_{\mu = 1}^n \left((\omega_B)_{\mu} - (\omega_A)_{\mu} \right) (\omega_A)_{\mu} x_{\mu}^2
\end{align}
\begin{align}
    \rho_B^{N,\text{QHO}} (\bm{x}) &= \prod_{\mu = 1}^n \sqrt{\frac{(\omega_B)_{\mu}}{(\omega_A)_{\mu}}} \,\, \rho_A^{N,\text{QHO}} \left( \Omega_A^{-1/2} \Omega_B^{1/2} \bm{x} \right)
\end{align}
where the matrix power applies to the main diagonal element-wise.
Still, $\rho_A^{N,\text{QHO}}$ is unknown. To extract analytical statements, we again apply the trick introduced above in Eq.~\ref{eq:double_difference}:
\begin{align}
    \label{eq:const_density_law}
    c_{\mu} &= \omega_{\mu} \int_{\mathbb{R}^n} d\bm{x}  \, x_{\mu}^2 \, \,\rho_{N,\text{QHO}} \\
    \label{eq:DeltaDeltaE}
    E &= \sum_{\mu = 1}^n c_{\mu} \, \omega_{\mu}
\end{align}
with~$c_{\mu}$ being constant with respect to~$\omega_{\mu}$.

\section{Discussion}
\label{sec:discusssions}

The transformations presented here, be they affine, of square-root type or with functions including~$||\bm{y}||_2$, connect a number of systems in $n$~dimensions and for multiparticle systems as well. Thus, if the solution to a Hamiltonian proves unfeasible, $\mathcal{K}[\Delta v](\bm{x})$~may provide systematic access to its energy and particle density behavior and aid with relative tools in finding eigen values \textit{and} eigen states. This point is worth repeating: finding a suitable parametrization~$\bm{r}(\lambda)$ of the coordinates of two Schrödinger equations awards immediate relative information about the energy spectra and the (excited) densities! Thus, it provides the study of chemical compound space with an analytical tool of navigation.

However, the constraints employed to find such a parametrization also restrict the set of compounds $A,B$ connected by this very parametrization. On one hand, this is a limitation of AIT and the corresponding chemical compound space. On the other hand, such restrictions on possible compounds equip materials design with a tool for localization. As the number of possible materials is colossal~\cite{ceder1998predicting}, pinpointing small subsets of compounds can be considered desirable.

Furthermore, the new formula of~$\mathcal{K}[\Delta v](\bm{x})$ in Eq.~\ref{eq:K_final_form} evades any problems of convergence inherent in its original derivation \cite{krug_DeltaE}. Although tested numerically for transmutations in atoms and molecules in Ref.~\onlinecite{rudorff2021arbitrarily}, here we need not worry about divergences in absence of an infinite series. However, AIT is a relative method; when calculating explicit values with Eq.~\ref{eq:AIT_setup}, its accuracy depends on the quality of the initial electron density~$\rho_A$ (and by extension, the integration algorithm).

It has not escaped our attention that homogeneous functions as potentials provide elegant connections between kinetic and potential energy in the virial theorem~\cite{cohentannoudji}. However, this relation is due to Euler's homogeneous function theorem (i.e. a consequence of the derivative behavior of homogeneous functions), while AIT obtains its parameter $k(\lambda)$ by taking the $(2+\nu)$-th root. Nonetheless, this is not the first time in which the mathematical properties of a homogeneous potential act in favor of a theorem.

% AIT stands out because firstly, it restricts the textbook statement of perturbation theory in quantum mechanics by Wigner that the $p$-th order wave function derivative is necessary to describe the $(2p+1)$-th energy derivative~\cite{Wigner_2n+1}, and secondly, it enables applications to arbitrary dimensions and external potentials for any affine transformation of the coordinates. 
% The remaining challenge, preventing the generic use of AIT to estimate changes between arbitrary systems, is the parameterization. 

Constraining AIT to homogeneous functions and scaling transformations ($A(\lambda) = k(\lambda)/k_A \, \mathbb{1}_n$) allowed a generalization of the three homogeneous examples (QHO, Hydrogen-like atom, Dirac well).
Whichever potential is used, Eqs.~\ref{eq:kernel_hom} and~\ref{eq:relation_hom_densities} allow statements about the solutions of Schrödinger's equations without the necessity to ever explicitly solve them.
In this regard, it becomes equivalent to coordinate scaling in DFT as described in Refs.~\onlinecite{levy_1991,nagy_1995,kuemmel_2008}.
% We need not be concerned with the accuracy of computational methods (= quality) or the sampling coverage for machine learning models, i.e. the enumeration of solutions (= quantity).
Note that the fraction of two homogeneous functions is again a homogeneous function. Consequently, one might refrain from Taylor-expanding a physical potential to model materials in arbitrary dimensions, and instead employ a Padé approximant which, in addition, exhibits better convergence properties and truncation error.

However, coordinate scaling in homogeneous potentials alone does not allow for additional constraints of the problem. Here, AIT stands out by including external constraints via Eq.~\ref{eq:ansatz_rho} like the known electric dipole moment in Eq.~\ref{eq:param_dipole}. In addition, we were able to treat the multiparticle system in Eq.~\ref{eq:Hamiltonian_multiQHO} where general solutions are difficult (or analytically impossible) to obtain, yet Eqs.~\ref{eq:const_density_law} and~\ref{eq:DeltaDeltaE} disclose information about the (excited) energies and conservation laws of the density. 
We are aware that many systems can be solved numerically to desired accuracy; however, we deem such analytical statements about the relationship between systems always preferable to numerical ones and thus, consider AIT to be an effective tool in navigating between systems.

% Finally, we want to emphasize the generality of the presented method: although quantum mechanics appears as an obvious application, AIT requires the density to appear explicitly after the Hellmann-Feynman theorem (which in turn applies to any eigen problem of any Hermitian operator in Hilbert space) to solve Eq.~\ref{eq:solve_for_N}. Consequently, any transformation of the coordinates in a Hermitian operator, that allows a solution of Eq.~\ref{eq:solve_for_N} or a similar constraint, bears an integral kernel like $\mathcal{K}$. A simplification like the one in Eq.~\ref{eq:HellmannFeynmanApplied} is not unique to the Schrödinger equation.

\section{Conclusion}
\label{sec:conclusion}

We have presented a simpler and more general kernel~$\mathcal{K}$ of AIT in $n$~dimensions, with a method to obtain parametrizations~$\bm{r}(\lambda)$, an application to homogeneous potentials of degree $\nu \neq -2$, and two instances of a multiparticle system (Eqs.~\ref{eq:homfunc_multi_Hamiltonian} and~\ref{eq:Hamiltonian_multiQHO}). AIT did not just predict the relative behavior of energy spectra, but relative densities as well. % Examples included one- and multiparticle systems like the quantum harmonic oscillator, the Hydrogen-like atom, and the Dirac well.
In doing so, we remedied issues discussed in a previous paper~\cite{krug_DeltaE} like the convergence problems of the kernel~$\mathcal{K}$ as a series, the unknown parametrization or any rigorous method of obtaining it, and an extension of AIT to analytical statements between systems with analytically unknown energies or densities.

Future work will deal with general parametrizations~$\bm{r}(\lambda)$ in $n$-dimensional systems. To solve Eq.~\ref{eq:ansatz_rho}, one might employ additional information about the initial electron density~$\rho_A$, similar to the presented known electric dipole moment in Eq.~\ref{eq:electric_dipole_moment}. Building on such additional constraints might provide an elegant path to connect not just the ground-state energies and densities of compounds $A$ and $B$, but all available excited states as well.

\section*{Supplementary Material}
The solution to the integrals in Eqs.~\ref{eq:integral_QHO} and~\ref{eq:integral_HL} and the explicit electron densities can be found in the~Supplemental Material.

\section*{Acknowledgements}
% Kieron: Z^2 part
% Danish and Florian, theo paper
% Dirk Andrae: holds generally for excitations
% Roi Baer: By virial theorem, AIT in monoatomics holds for T and V separately
We acknowledge discussions with Dirk Andrae, Kieron Burke, Florian Bley and Danish Khan.
O.A.v.L. has received funding from the European Research Council (ERC) under the European Union’s Horizon 2020 research and innovation programme (grant agreement No. 772834). 
This research was undertaken thanks in part to funding provided to the University of Toronto's Acceleration Consortium from the Canada First Research Excellence Fund, grant number: CFREF-2022-00042. O.A.v.L. has received support as the Ed Clark Chair of Advanced Materials and as a Canada CIFAR AI Chair.

\section*{Author contributions}
\textbf{S.L.K.} conceptualization (lead), formal analysis (lead), investigation (lead), methodology (lead), writing - original draft (lead), writing - review \& editing (equal).
\textbf{O.A.v.L.} conceptualization (supporting), formal analysis (supporting), investigation (supporting), methodology (supporting), funding acquisition, project administration, resources, supervision (lead), writing - review \& editing (equal)

\section*{Conflict of Interest}
The authors have no conflicts to disclose.
% \section*{Author Contributions}
% Contributions are defined via the NISO standard:
% https://www.niso.org/publications/z39104-2022-credit
% \textbf{Simon León Krug}: conceptualization (lead), data curation, formal analysis (lead), investigation (lead), methodology (lead), software, visualization (equal), writing - original draft (lead), writing - review \& editing (supporting).
% \textbf{O.~Anatole von~Lilienfeld}: conceptualization (supporting), formal analysis (supporting), investigation (supporting), methodology (supporting), funding acquisition, project administration, resources, supervision (lead), visualization (equal), writing - review \& editing (lead).
All authors read and approved the final manuscript.

% \section*{Data and code availability}
% % IN SECOND VERSION: MOVE SCRIPT OF FIG 1 AND FIG 2 TO ZENODO, PROVIDE LINK
% % IF APPLICABLE, ADD ALSO THE FIG 3 WHERE QHO AND MORSE ARE PLOTTED AGAIN BUT WIHTOUT CONVERGENCE ISSUES
% % Fix abstratc in arxiv version Z -> Z_B!

% The code that produces the findings of this study, in specific the comparisons of the quantum harmonic oscillator and the Morse potential, is openly available on GitHub under \url{https://github.com/SimonLeonKrug/pyalchemy}.

\bibliography{refs.bib}{}
\bibliographystyle{unsrt}

%\begin{comment}

\clearpage
%\appendix
\onecolumngrid
\setcounter{section}{0}

\begin{center}
    \large \textbf{\papertitle}\\
    \large \textbf{ ---  Supplemental Information  ---}\\
    \vspace{\baselineskip}
    \normalsize Simon León Krug,$^{1}$ and O. Anatole von Lilienfeld$^{1,2,3,4,5,6,7}$\\
    \vspace{0.5\baselineskip}
    \small \textit{
    $^{1)}$Machine Learning Group, Technische Universität Berlin,
    10587 Berlin, Germany\\
    $^{2)}$Berlin Institute for the Foundations of Learning and Data, 10587 Berlin, Germany\\
    $^{3)}$Chemical Physics Theory Group, Department of Chemistry, University of Toronto, St. George Campus, Toronto, ON, Canada\\
    $^{4)}$Department of Materials Science and Engineering, University of Toronto, St. George Campus, Toronto, ON, Canada\\
    $^{5)}$Vector Institute for Artificial Intelligence, Toronto, ON, Canada\\
    $^{6)}$Department of Physics, University of Toronto, St. George Campus, Toronto, ON, Canada\\
    $^{7)}$Acceleration Consortium, University of Toronto, Toronto, ON, Canada\\
    }
    \small (Dated: \today)
\end{center}

\section{Integrals of the electron density}

\subsection{The quantum harmonic oscillator}
\begin{align}
    \label{eq:QHO_x_squared}
    \int\limits_{-\infty}^{+\infty} dx \, x^2 \rho_A^{\text{QHO}}(x)
    &= \frac{1}{\omega_A} \left(n + \frac{1}{2} \right)
\end{align}
where the electron density of the quantum harmonic oscillator is given as~\cite{cohentannoudji}:
\begin{align}
    \rho_A^{\text{QHO}}(x) &= \frac{1}{2^n \, n!} \sqrt{\frac{\omega_A}{\pi}} e^{-\omega_A x^2} H_n^2 (\sqrt{\omega_A}x)
\end{align}
\\ \\
PROOF:
\begin{align}
    \int\limits_{-\infty}^{+\infty} dx \, x^2 \rho_A^{\text{QHO}}(x) &= \frac{1}{2^n \, n!} \sqrt{\frac{\omega_A}{\pi}} \int\limits_{-\infty}^{+\infty} dx \, x^2 e^{-\omega_A x^2} H_n^2 (\sqrt{\omega_A}x) \\
    &= \frac{1}{2^n \, n! \sqrt{\pi} \, \omega_A} \int\limits_{-\infty}^{+\infty} dy \, y^2 e^{-y^2} H_n^2 (y)
\end{align}
Now use the recurrence relation
\begin{align}
    y \, H_n(y) = \frac{H_{n+1}(y) + 2nH_{n-1}(y)}{2}
\end{align}
and the orthogonality relation
\begin{align}
    \int\limits_{-\infty}^{+\infty} dy \, e^{-y^2} H_n (y) \, H_m (y) = \sqrt{\pi} \, 2^n \, n! \delta_{nm}
\end{align}
such that:
\begin{align}
    \int\limits_{-\infty}^{+\infty} dx \, x^2 \rho_A^{QHO}(x) &= \frac{1}{2^n \, n! \sqrt{\pi} \, \omega_A} \int\limits_{-\infty}^{+\infty} dy \, y^2 e^{-y^2} H_n^2 (y) \\
    &= \frac{1}{2^n \, n! \sqrt{\pi} \, \omega_A} \int\limits_{-\infty}^{+\infty} dy \, e^{-y^2} \left( \frac{H_{n+1}(y) + 2nH_{n-1}(y)}{2} \right)^2 \\
    &= \frac{1}{2^{n+2} \, n! \sqrt{\pi} \, \omega_A} \int\limits_{-\infty}^{+\infty} dy \, e^{-y^2} \left( H^2_{n+1}(y) + 4n^2 H^2_{n-1}(y)\right) \\
    &= \frac{1}{2^{n+2} \, n! \sqrt{\pi} \, \omega_A} \sqrt{\pi} \left( 2^{n+1} (n+1)! + 4n^2 \, 2^{n-1} (n-1)!\right)\\
    &= \frac{1}{2\omega_A} \left(n+1 + n\right) \\
    &= \frac{1}{\omega_A} \left(n + \frac{1}{2} \right)
\end{align}

\subsection{The Hydrogen-like atom}
\begin{align}
    \label{eq:hydlike_over_x}
    \int_{\mathbb{R}^3} d\bm{x} \,\frac{\rho_A^{\text{HL}}(\bm{x})}{||\bm{x}||_2}
    &= \frac{Z_A}{n^2}
\end{align}
where the electron density of the Hydrogen-like atom is given as~\cite{cohentannoudji}:
\begin{align}
    \label{eq:density_HL}
    \rho_A^{\text{HL}}(\bm{x}) &= 
    \left( \frac{2Z_A}{n} \right)^3 \frac{(n-l-1)!}{2n (n+l)!} \left( \frac{2Z_A ||\bm{x}||_2}{n} \right)^{2l} \left( L_{n-l-1}^{(2l+1)}\left( \frac{2Z_A||\bm{x}||_2}{n}\right) \right)^2 \exp{\left(- \frac{2Z_A||\bm{x}||_2}{n}\right)} \, |Y_{lm}|^2
\end{align}
with generalized Laguerre polynomials $L^{(\alpha)}_{m}$ and spherical harmonics $Y_{lm}$.
\\ \\
PROOF:\\ \\
Note the orthogonality relation of the generalized Laguerre polynomials:
\begin{align}
    \int_0^{\infty} d\nu \,  e^{-\nu} \nu^{\alpha} L_{m}^{(\alpha)}\left( \nu \right) L_{m'}^{(\alpha)}\left( \nu \right)  = \frac{\Gamma (m+\alpha+1)}{m!} \delta_{m,m'}
\end{align}
and the orthogonality relation of the spherical harmonics:
\begin{align}
    \int_{S^2} d\Omega \, Y_{lm}  Y^*_{l'm'} = \delta_{ll'} \delta_{mm'}
\end{align}

First, execute the angular integration, then substitute $\nu = 2Z_A ||\bm{x}||_2/n$:
\begin{align}
    \int_{\mathbb{R}^3} d\bm{x} \, \frac{\rho_A^{\text{HL}}(\bm{x})}{||\bm{x}||_2}
    &= \int_0^{\infty} d||\bm{x}||_2 \, ||\bm{x}||_2 \left( \frac{2Z_A}{n} \right)^3 \frac{(n-l-1)!}{2n (n+l)!} \left( \frac{2Z_A ||\bm{x}||_2}{n} \right)^{2l} \left( L_{n-l-1}^{(2l+1)}\left( \frac{2Z_A||\bm{x}||_2}{n}\right) \right)^2 \exp{\left(- \frac{2Z_A||\bm{x}||_2}{n}\right)} \\
    &= \frac{2Z_A}{n} \frac{(n-l-1)!}{2n (n+l)!} \int_0^{\infty} d\nu \,  e^{-\nu} \nu^{2l+1} \left( L_{n-l-1}^{(2l+1)}\left( \nu \right) \right)^2 \\
    &= \frac{2Z_A}{n} \frac{(n-l-1)!}{2n (n+l)!} \frac{(2l+1+n-l-1)!}{(n-l-1)!} \\
    &= \frac{Z_A}{n^2}
\end{align}

% \section{Software}
% Software for the purpose of data generation (i.e. math libraries, integration algorithms, numerical tools) are provided by the \texttt{Python}-packages \texttt{NumPy}\cite{numpy} and \texttt{SciPy}\cite{scipy}. Visualizations were created using \texttt{Matplotlib}\cite{matplotlib}. Calculations were often performed or cross-checked using \texttt{Mathematica}\cite{mathematica}.

\end{document}